\documentstyle [12pt] {article}

\title{
The theory of the scattering-induced feeding-in in bent crystals
}

\author{Valery Biryukov\thanks{Electronic address:  biryukov@mx.ihep.su  }
\\ Institute for High Energy Physics, \\
 Protvino, 142284 Moscow Region, Russia }

\date{}
\begin{document}
\normalsize
\pagestyle{myheadings}
\markright{\em Physics Letters A {\bf 205}, 340--343 (1995)}

\maketitle

\leftline{Published in: \em Physics Letters A {\bf 205}, 340--343 (1995)}

\begin{abstract}
\normalsize
\baselineskip=15pt
An analytical theory for the efficiency of scattering-induced transitions
from a random to a channeled state (feed-in) in bent crystals is derived.
The predictions from the theory are in good agreement with
experiment and Monte Carlo simulations.
\end{abstract}

\normalsize
\baselineskip=15pt

\section{Introduction}

Steering of particle beams by means of channeling
in bent crystals \cite{tsyg} is rapidly evolving technique
for particle accelerators \cite{ufn}.
The particle motion in a crystal is influenced by a series of collisions
with the crystal constituents, which
may cause particle transitions between channeled and random states.
For any trajectory of a particle in a crystal a time-reversed
trajectory is possible.
The starting point of a trajectory
becomes the final one, and vice versa.
This leads  to the idea of { reversibility}
of transition processes \cite{li}.
In the depth of a crystal,  besides the particles
leaving the channeling mode (dechanneling or {\em feeding out}),
there may be particles entering the channeling mode
(respectively, {\em feeding in}, also known as {\em volume capture}).
The mechanisms, responsible for these two opposite processes,
are essentially the same.
Here we only discuss the scattering-induced transitions.
Another mechanism of feeding-in,
namely the centrifugal effects, is discussed elsewhere \cite{feed2}.

The feed-out process is well described by both the analytical theory
and the computer simulations (Ref. \cite{ufn} and references therein).
The feed-in has been studied numerically,
while the analytical theory was missing.
Here we derive an explicit formula for the
efficiency of the scattering-induced feeding-in in bent crystals.

\section{Feed-in efficiency in bent crystals}
\label{der2}

Let us consider a beam with a uniform angular distribution,
1/2$\Phi$, incident on a bent crystal with  curvature 1/$R$.
The fraction of beam channeled is a function
of the crystal depth $z$:
\begin{equation}	\label{benaa}
f(z)
 = \frac{2x_c}{d_p}
    \frac{\pi}{4}
    \frac{\theta_{c}}{\Phi}
    A_B(pv/R)
        F_D(z)
\end{equation}
where $A_B(pv/R)$ describes the reduction of the bent-crystal acceptance
for a particle with momentum $p$ and velocity $v$;
$\theta_c$ and $x_c$ are the critical angle
and transverse position, respectively; $d_p$ is the interplanar spacing.
The factor $\pi /4$ is exact in a harmonic case;
for the realistic potential it should be replaced with $\approx$0.8.
The $F_D(z)$ describes the feed-out of the channeled particles
due to scattering;
usually one writes $F_D(z)$ as $\exp (-z/L_D)$
($L_D$ is the dechanneling length),
and so $F_D'(z)=-F_D(z)/L_D$.

The number of particles dechanneled over the length
$\delta z$ equals
\begin{equation}	\label{b2a}
-\frac{df(z)}{dz}
 = -\frac{2x_c}{d_p}
    \frac{\pi}{4}
    \frac{\theta_{c}}{\Phi}
    A_B(pv/R)
    F_D'(z)
 = \frac{2x_c}{d_p}
    \frac{\pi}{4}
    \frac{\theta_{c}}{\Phi}
    A_B(pv/R)
    \frac{F_D(z) }{L_D}
\end{equation}
The particles dechanneled over $dz$ are exiting in the angular range
$d\theta =dz/R$. Therefore the angular distribution
downstream of the crystal is
\begin{equation}	\label{b3a}
\frac{df}{d\theta}
 = R\frac{df(z)}{dz}
 = \frac{2x_c}{d_p}
    \frac{\pi}{4}
    \frac{\theta_{c}}{\Phi}
    A_B(pv/R)
      \frac{R}{L_D} F_D(z)
\end{equation}
We don't take into account a small extra spreading,
$\pm\theta_c$, of the dechanneled particles.
Therefore, for big $R$ (when $L_D/R$ is comparable to $\theta_c$),
Eq. (\ref{b3a}) overestimates the phase density.

Let us now consider the same beam
incident on the same crystal in the {\em reverse} direction.
Now the particles with the upstream parameters ($x_i, \theta_i$)
equal to the downstream parameters ($x_f, \theta_f$)
of the dechanneled particles from the preceding case,
are captured along the same (reversed) trajectories.
By consideration, the number of particles
which have experienced  transitions {\em from}  the channeled states
in the former case,
is equal to the number of transitions {\em to} the channeled states
in the latter case (as the trajectories are the same).

Therefore the number of particles captured from the interval $d\theta$
and then transmitted in the channeled states
(over the length $z$) to the crystal face,
is given by Eq.\ (\ref{b3a}).
We write this number as $w_SF_D(z)$, the product of the
capture probability and the transmission factor.
The transmission factor $F_D(z)$,
for particles channeled in the same states,
is the same irrespective of the direction of motion.
Normalizing Eq.\ (\ref{b3a}) to the number of
particles incident on the crystal
in this angular range, $d\theta /2\Phi$,
one obtains the capture probability:
\begin{equation}	\label{wb2a}
w_S = 2\Phi\frac{df(z)}{d\theta} \frac{1}{F_D(z)}
 = \frac{\pi x_c}{d_p}
    \frac{R\theta_c}{L_D(pv/R)}
    A_B(pv/R)
\end{equation}
For a harmonic potential, $A_B$$=$$(1-R_c/R)^2$
where $R_c$ is the critical radius;
at the same time, $L_D$ is reduced in a bent crystal by the same factor
$(1-R_c/R)^2$ relative to $L_D$ in unbent crystal \cite{ufn}.
For a realistic potential the ratio of the two factors is $\approx$1.
Hence we can omit $A_B$ in (\ref{wb2a}) and imply $L_D$ value
for a straight crystal; another simplification is $x_c/d_p\approx$1/2.
One obtains:
\begin{equation}	\label{ws}
w_S  = \; \frac{\pi x_c}{d_p} \frac{R\theta_c}{L_D} \;
     \approx  \;  \frac{\pi}{2} \frac{R\theta_c}{L_D}
\end{equation}
The fact that $w_S$ should be of the order of $R\theta_c/L_D$
has been earlier found from a simple qualitative consideration
\cite{bir-b73}.
To see the explicit dependence of Eq. (\ref{ws})
on the properties of crystal and the particle energy,
one can use the formulas for $\theta_c$ \cite{li}
\begin{equation}	\label{thetac}
\theta_c = \left( \frac{4\pi Nd_pZ_iZe^2a_{TF}}{pv}\right)^{1/2} \; ,
\end{equation}
and for $L_D$ \cite{bir94}
         \begin{equation}              \label{ld}
      L_D=      \frac{256}{9\pi^2}\cdot
             \frac{pv}{\ln (2m_ec^2\gamma/I)-1}
              \cdot \frac{a_{TF}d_p}{Z_ie^2} \; ,
            \end{equation}
where
$N$ is the volume density of atoms,
$Z$ the atomic number, $Z_ie$ the particle charge,
$a_{TF}$ the Thomas-Fermi screening distance,
$m_e$ the electron rest mass,
$I\simeq 16Z^{0.9}$ eV is the ionization potential,
$\gamma$ the particle Lorentz factor.
Eq. (\ref{ws}) takes the form:
\begin{equation}	\label{wss}
w_S =  \frac{9\pi^{7/2}}{256} \left(\frac{Z_i}{pv}\right)^{3/2}
         \cdot R \cdot
        \frac{N^{1/2}Z^{1/2}e^3}{d_p^{1/2}a_{TF}^{1/2}}
	 \Bigl(\ln (2m_ec^2\gamma/I)-1\Bigr)   \;  .
\end{equation}
It should be mentioned that the length $L_D$ for dechanneling from
the 'stable states' is well defined
and has been measured in many experiments,
in good agreement with Eq. (\ref{ld}) \cite{bir94}.
Therefore, the rate of feeding-in {\em to}
the same states is an equally-well defined quantity.

The feed-in rate has been studied in bent crystals
experimentally at up to 70 GeV \cite{liaf,ches90}.
It was found that $w_S$ is proportional to $R$ and
to $p^{-3/2}$, which is in perfect agreement with Eq. (\ref{ws}).
The Table 1 shows the probabilities of capture into the 'stable states'
(which dechannel by exponential law) measured in the experiment
at 70 GeV \cite{ches90}, found in the Monte Carlo simulation \cite{bi95},
and calculated by Eq. (\ref{ws}).
There is agreement within $\simeq$10--20 \%, i.e. within the errors of
the experiment and simulation.

\begin{table}
\caption{
The probability (in \%) of the feed-in into ''stable states'',
for 70 GeV proton in Si crystal bent with $R$=3 m,
from the experiment,
simulation, and Eq. (5). }
\begin{center}
\begin{tabular}{||c||c|c|c||}
\hline
{  } & { } & { } & { }  \\
{Crystal} & {Eq. (5) } & {Simulation} &
{Experiment } \\
{  } & { } & { } & { }  \\
\hline
\hline
{ } & { } & { } & { }  \\
{ 111 } & {0.20} & {0.17$\pm$0.02} & {0.23} \\
{ } & { } & { } & { }  \\
                      \hline
{ } & { } & { } & { }  \\
{ 110 } & {0.22} & {0.23$\pm$0.02} & { -- } \\
{ } & { } & { } &
{ }  \\
                      \hline
                \hline
\end{tabular}
\end{center}
\label{mctab1}
\end{table}

Few remarks on Eq. (\ref{ws}).
The feed-out rate 1/$L_D$ at a very small depth in crystal
is larger due to the 'short-lived' states with high transverse energies.
Then we find from our consideration that the feed-in rate
$w_S\sim R\theta_c/L_D$ {\em to} such states is respectively larger,
which is qualitatively obvious.

With $R$ increase, Eq. (\ref{ws}) may become $>1$.
The reason is the neglect of the $\pm\theta_c$ spreading in Eq. (\ref{b3a}).
If one sums quadratically the divergences,
$\pm\theta_c$ and that from (\ref{b3a}),
then the capture probability does not exceed 1.
However, notice that $w_S$ is as small as $\sim$1 \%
in the experimental practice.

The reversibility relation (\ref{ws}) is valid
irrespective of the mechanism of feeding-out (-in).
In the presence of {\em dislocations}, $L_D$ is small
and there should be a high rate of feeding-in,
in accordance with Eq. (\ref{ws}).
This relation was observed indeed in the experiment \cite{ches}
with dislocation-contaminated crystal of germanium and the 70 GeV
proton beam.

\section{Applications}

Although the feed-in processes cannot increase
the beam bending efficiency,
they contribute to a wide angular acceptance ($\gg\theta_c$)
crystal deflector. The bending efficiency
of such deflector
can be {\em designed}, and be varied in a broad range of values.
Moreover, this efficiency is independent of the incident beam divergence
(in contrast to the regular case of the entry-face capture).
This may be valuable,
when the beam attenuation, stability, and low background are
important issues.
Obviously, the secondary particles produced in collisions with crystal
nuclei can be captured into the channeling mode (from the crystal bulk)
through the feed-in processes only.
This may be applicable, e.g., in some ideas of crystal application
for experiments in particle physics \cite{relchan}.

\end{document}